\documentclass[english,english,showpacs,prb,preprint,tightenlines,amsmath,floatfix]{revtex4}
\usepackage[T1]{fontenc}
\usepackage[latin1]{inputenc}
\usepackage{array}
\usepackage{graphicx}
\usepackage{longtable}

\providecommand{\tabularnewline}{\\}

\makeatletter

\usepackage{float}

\makeatletter

\makeatletter



\makeatother

\makeatother

\makeatother

\usepackage{babel}

\usepackage{babel}

\usepackage{babel}

\usepackage{babel}

\usepackage{babel}

\makeatother

\usepackage{babel}

\usepackage{babel}

\begin{document}

\title{Large-scale correlated study of excited state absorptions in naphthalene
and anthracene}

\author{Priya Sony}

\author{Alok Shukla}

\affiliation{Department of Physics, Indian Institute of Technology Bombay, Powai,
Mumbai 400076, INDIA.}
\begin{abstract}
In this paper, we report theoretical calculations of the photo-induced
absorption (PA) spectrum of naphthalene and anthracene, with the aim
of understanding those excited states, which are invisible in the
linear optical absorption. The excited state absorption spectra are
computed from the $1B_{2u}^{+}$ and the $1B_{3u}^{+}$ states, and
a detailed analysis of the many-body character of the states contributing
to various peaks in the spectra is presented. The calculations are
performed using the Pariser-Parr-Pople (PPP) Hamiltonian, along with
the full configuration interaction (FCI) technique. The role of Coulomb
parameters used in the PPP Hamiltonian is examined by considering
standard Ohno parameters, as well as a screened set of parameters.
The results of our calculations are extensively compared with the
experimental data where available, and very good agreement has been
obtained. Moreover, our calculations predict the presence of high
intensity features which, to the best of our knowledge have not been
explored earlier. We also present concrete predictions on the polarization
properties of the PA spectrum, which can be verified in experiments
performed on oriented samples. 
\end{abstract}

\pacs{78.30.Jw, 78.20.Bh, 42.65.-k }

\maketitle

\section{Introduction}

Conjugated polymers have already replaced the traditional inorganic
materials and emerged as popular choice for manufacturing opto-electronic
devices.\citet{baldo2000,forrest2004,gershenson2006,malliaras2005}
The inter-molecular localized and the intra-molecular delocalized
$\pi-$electrons are credited for the interesting linear and nonlinear
optical properties in these materials. The large nonlinear optical
(NLO) response in such materials is favorable for various device applications
such as optical switching, computing, and electro-opto modulation.\citet{bredas1994,Eason1993,zyss1994}
These useful applications have motivated scientists to study the excited
states which are important for understanding the nonlinear optics
of these materials. For example, in centrosymmetric molecules, states
can be classified as even parity (\emph{gerade}) and odd parity (\emph{ungerade})
states. From the ground state, using linear optical spectroscopy,
one can probe the low-lying odd parity excited states,\citet{sony-acene-lo}
while the even parity states stay invisible. Photo-induced absorption
(PA) technique gives an opportunity to investigate these even parity
states, which contribute to the nonlinear optical processes in these
materials.

\begin{figure}
\includegraphics[width=8cm]{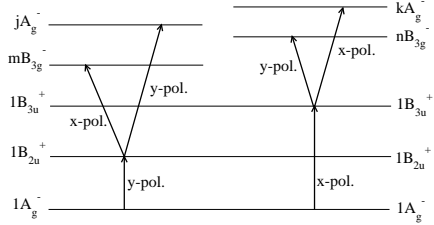}

\caption{Schematic diagram of the essential states involved in the PA in oligoacenes,
and their polarization characteristics. The arrows connecting two
states imply optical absorption, with polarization directions stated
next to them. Location of states is not up to scale.\label{fig:essential states}}

\end{figure}

Among various conjugated polymers, oligo-acenes, due to their interesting
electronic and optical properties, have come up as potential candidates
for possible applications in the field of nonlinear optics.\citet{dimitrakopoulos2002,muller2005}
Because of the alternant nature of these molecules, different states
in addition to the point group symmetry ($D_{2h}$) can also be classified
according to their particle-hole parity ($\pm$). According to normal
convention, we assign the ground state ${1A}_{g}$ a negative ($-$)
particle-hole parity. As per the selection rules of $D_{2h}$ point
group symmetry, for the ground-state linear optical absorption in
oriented samples of oligo-acenes, two polarization channels are possible.
Absorption of $y$-polarized photons leads the system to the $B_{2u}^{+}$
manifold, while the $x$-polarized photons cause transitions to the
$B_{3u}^{+}$ states.\citet{sony-acene-lo} Same selection rules,
therefore, also offer both these polarization possibilities for the
excited state absorptions from the $B_{2u}^{+}$ and $B_{3u}^{+}$
states. Thus, from the $B_{2u}^{+}$ type states, using $x/y$-polarized
photons, $B_{3g}^{-}$/$A_{g}^{-}$-type excited states can be reached,
while from the $B_{3u}^{+}$ states, using $x/y$-polarized photons,
$A_{g}^{-}/$$B_{3g}^{-}$ can be obtained (\emph{cf.} Fig. (\ref{fig:essential states})).
Here, the noteworthy point is the reverse symmetry-polarization relationship
for the two types of manifolds. In the 70's, Bergman and Jortner\citet{bergman-2A,bergman-3A}
studied, these even parity excited states ($A_{g}^{-}$ \& $B_{3g}^{-}$)
using the two-photon absorption (TPA) spectroscopy. They measured
the TPA spectra of naphthalene and anthracene in crystalline as well
as in solution mode. Later, Dick and Hohlneicher measured the spectra
over a wide range, between $3.59$~eV to $6.20$~eV for the naphthalene\citet{dick-2A}
and anthracene\citet{dick-3A} molecules in solution. As far as the
theory is concerned, the calculations have been performed using various
methodologies like SDCI,\citet{tavan1979} CASSCF,\citet{hashimoto-2A,kawashima-3A-4A}
MRMP,\citet{hashimoto-2A,kawashima-3A-4A} LCAO-MO,\citet{pariser-2A-5A}
exact PPP,\citet{ramasesha-soos91,ramasesha-3A} CNDO/S-CI,\cite{Swiderek1990}
and CNDO-CI\citet{hohlneicher-2A-3A} on the molecular phases of these
oligomers. To the best of our knowledge, none of the above mentioned
calculations have reported the photo-induced absorption spectrum of
any of these materials. Moreover, most of the calculations are restricted
to the low-order treatment of electron-correlation effects. Therefore,
in this paper, we present a large scale correlated study of photo-induced
absorption spectra of the first two members of acene family, namely,
naphthalene and anthracene (\emph{c.f.} Fig. (\ref{fig-acene})).
The PA spectra from the $1B_{2u}^{+}$ and the $1B_{3u}^{+}$ states
are computed, and a detailed analysis of the many-body character of
the important excited states contributing to the spectra is performed.
The calculations are done using the Pariser-Parr-Pople (PPP) model
Hamiltonian along with full configuration interaction (FCI) technique.
The standard,\citet{ohno} as well as the screened parameters\citet{chandross}
are used in the PPP model Hamiltonian, and the results from both sets
of calculations are compared with the experiments so as to analyze
the role of Coulomb parameters. Low energy features of our calculated
PA spectra match very well with the peaks observed in the two photon
spectroscopy (TPS) based experiments.\cite{dick-2A,dick-3A,mikami-2A,bergman-2A,bergman-3A}
Moreover, our calculations predict the presence of intense absorption
features in the higher energy range of the PA spectra, along with
their polarization characteristics. Particularly, for the case of
anthracene novel high intensity features well below the ionization
threshold should be detectable in future experiments. Predicted polarizations
properties of various features can also be tested in measurements
performed on oriented samples.

The remainder of this paper is organized as follows. In Sec. \ref{theory}
we briefly review the theoretical methodology adopted in this work.
In Sec. \ref{sec:Results} we present and discuss our results for
the photo-induced absorption from the $1B_{2u}^{+}$ and the $1B_{3u}^{+}$
~excited states. A comparison is made with the experimental and the
other available theoretical results. Finally, in Sec. \ref{sec:CONCLUSIONS}
we summarize our conclusions and provide prospects for future work.

\section{Theory}

\label{theory}

\begin{figure}
\includegraphics[width=8cm]{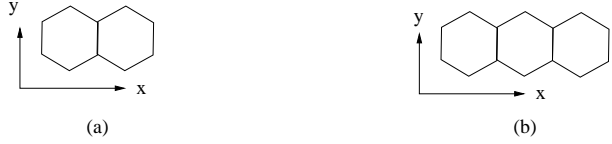}

\caption{Schematic drawings of (a) naphthalene and (b) anthracene, considered
in this work\label{fig-acene}}

\end{figure}

The schematic structures of naphthalene and anthracene are shown in
Fig.\ref{fig-acene}. Both the molecules are assumed to lie in the
$xy$-plane with the conjugation direction taken to be along the $x$-axis.
The carbon-carbon bond length has been fixed at $1.4$ \AA , and
all bond angles have been taken to be 120$^{o}$. The reason of choosing
this symmetric geometry, as against various other possibilities has
already been discussed in our earlier paper.\citet{sony-acene-lo}
Note that these structures can also be seen as two polyene chains
of suitable lengths, coupled together along the $y$-direction.

The correlated calculations are performed using the PPP model Hamiltonian,
which can be written as

\begin{equation}
H=H_{C_{1}}+H_{C_{2}}+H_{C_{1}C_{2}}+H_{ee},\label{eq-ham}\end{equation}
 where $H_{C_{1}}$ and $H_{C_{2}}$ are the one-electron Hamiltonians
for the carbon atoms located on the upper and the lower polyene like
chains, respectively. $H_{C_{1}C_{2}}$ is the one-electron hopping
between the two chains, and H$_{ee}$ depicts the electron-electron
repulsion. The individual terms can now be written as,\begin{subequations}
\label{allequations}

\begin{equation}
H_{C_{1}}=-t_{0}\sum_{\langle k,k'\rangle}B_{k,k'},\label{eq-h1}\end{equation}
 \begin{equation}
H_{C_{2}}=-t_{0}\sum_{\langle\mu,\nu\rangle}B_{\mu,\nu},\label{eq-h2}\end{equation}
 and

\begin{equation}
H_{C_{1}C_{2}}=-t_{\perp}\sum_{\langle k,\mu\rangle}B_{k,\mu}.\label{eq-h3}\end{equation}
 \end{subequations}

\begin{eqnarray}
H_{ee} & = & U\sum_{i}n_{i\uparrow}n_{i\downarrow}+\frac{1}{2}\sum_{i\neq j}V_{i,j}(n_{i}-1)(n_{j}-1)\textrm{}\label{eq:hee}\end{eqnarray}
 In the equation above, $k$, $k'$ are carbon atoms on the upper
polyene chain, $\mu,\nu$ are carbon atoms located on the lower polyene
chain, while $i$ and $j$ represent all the atoms of the oligomer.
Symbol $\langle...\rangle$ implies nearest neighbors, and $B_{i,j}=\sum_{\sigma}(c_{i,\sigma}^{\dagger}c_{j,\sigma}+h.c.)$.
Matrix elements $t_{0}$, and $t_{\perp}$ depict one-electron hops.
As far as the values of the hopping matrix elements are concerned,
we took $t_{0}=2.4$ eV for both intracell and intercell hopping,
and $t_{\perp}=t_{0}$ consistent with the undimerized ground state
for polyacene argued by Raghu \emph{et al}.\citet{Ramasesha-1}

The Coulomb interactions are parametrized according to the Ohno relationship,\citet{ohno}
\begin{equation}
V_{i,j}=U/\kappa_{i,j}(1+0.6117R_{i,j}^{2})^{1/2}\;\mbox{,}\label{eq-ohno}\end{equation}

where, $\kappa_{i,j}$ depicts the dielectric constant of the system
which can simulate the effects of screening, $U$ is the on-site repulsion
term, and $R_{i,j}$ is the distance in \AA ~ between the $i$th
carbon and the $j$th carbon. In the present work, we have performed
calculations using {}``standard parameters''\cite{ohno} with $U=11.13$
eV and $\kappa_{i,j}=1.0$, as well as {}``screened parameters''\cite{chandross}
with $U=8.0$ eV and $\kappa_{i,j}=2.0$ ($i\neq j$) and $\kappa_{i,i}=1.0$.
The screened parameters employed here were devised by Chandross and
Mazumdar\cite{chandross} with the aim of accounting for the interchain
screening effects in phenylene based polymers. Our motive behind using
these parameters is precisely the same.

The starting point of the correlated calculations for the molecules
are the restricted Hartree-Fock (HF) calculations, using the PPP Hamiltonian.
The many-body effects beyond HF are computed using the full configuration
interaction (FCI) method. Details of this CI-based many-body procedures
have been presented in our earlier works.\citet{shukla2,shukla-ppv,shukla-tpa,shukla-thg}
From the CI calculations, we obtain the eigenfunctions and eigenvalues
corresponding to the correlated ground and excited states of examined
molecules. Using these eigenfunctions the dipole matrix elements amongst
various excited states are computed. These dipole matrix elements,
along with the energies of the excited states are, in turn, utilized
to calculate various PA spectra. The number of excited states computed
for each symmetry manifold is quite large, ranging typically from
$70$--$100$, allowing us to probe a broad energy range in the PA
spectrum.

\section{Results and Discussion}

\label{sec:Results}
\begin{figure*}
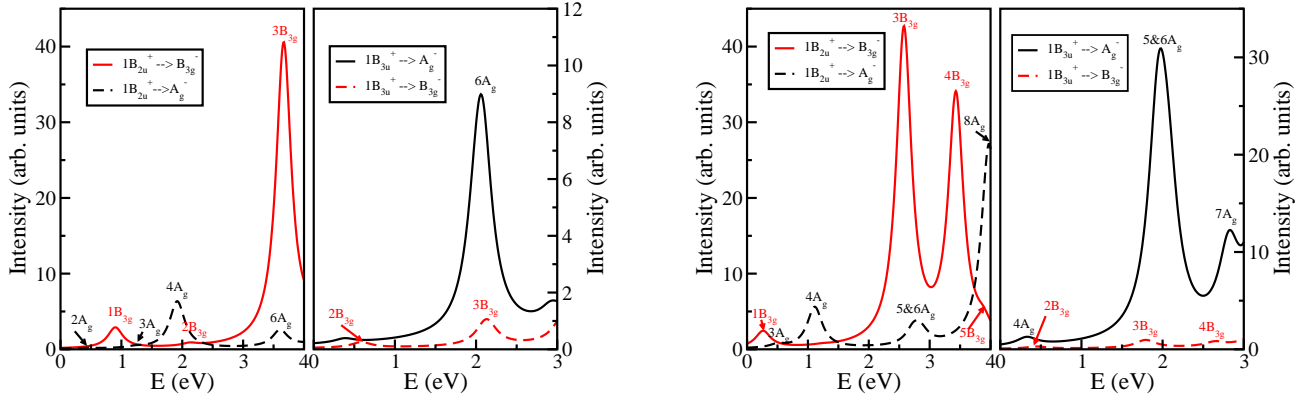

\hspace{-0.7cm}\includegraphics[width=8cm]{fig3a}\hspace{1.0cm}
\includegraphics[width=8cm]{fig3b}

\caption{(color online) Photo-induced spectra of naphthalene computed using
standard parameters (left panel) and screened parameters (right panel)
in the PPP model Hamiltonian. The left side of both the spectrum depicts
the PA from the $1B_{2u}^{+}$ state, while right side shows it from
the $1B_{3u}^{+}$ state. The $x$($y$)-polarized transitions are
shown using solid (dashed) curves.\label{fig:acene2_pa}}

\end{figure*}

In this section, we discuss our FCI results on
the PA spectrum of naphthalene and anthracene. Fig. (\ref{fig:acene2_pa}
and \ref{fig:acene3_pa}) presents the PA spectra of naphthalene and
anthracene, respectively. The spectra are computed using the standard
(left panel) as well as the screened parameters (right panel) in the
PPP Hamiltonian. For both $1B_{2u}^{+}$ and $1B_{3u}^{+}$ excited
state absorption, we present the $x$- and $y$-polarized spectra
separately, so as to facilitate comparison with experiments performed
on oriented samples. The wave functions of the excited states contributing
to various peaks in spectra are tabulated in tables (\textit{{cf.}}
Ref. \cite{epaps}). Since, the calculations are performed using the
FCI approach, they are exact within the model chosen and cannot be
improved. Therefore, any discrepancy which these results may exhibit
with respect to the experiments is a reflection of the limitations
of the model or the parameters used, and not that of the correlation
approach. At this point, we would like to note that Ramasesha and
coworkers have also presented FCI results on naphthalene\cite{Ramasesha-2,ramasesha-soos91}
and anthracene.\cite{ramasesha-3A,Ramasesha-2} However, for the case
of naphthalene they only presented the energies of some of the low-lying
states, but did not calculate transition dipoles. While, for anthracene
they computed the transition dipoles among a few even and odd parity
states, but the PA spectrum was not probed. Moreover, the energy range
of the states computed by us extends higher than the calculations
performed earlier, and therefore, can be used to interpret future
experiments.

In our earlier paper,\cite{sony-acene-lo} we showed that experiments
as well as calculations suggest that for both the oligomers $1B_{3u}^{+}$
state is located at least 1.5 eV higher than $1B_{2u}^{+}$ state.
Therefore if in a PA experiment, the pump is tuned close to $1B_{2u}^{+}$
energy, using an appropriate probe beam one can investigate higher
excited states which are dipole connected to $1B_{2u}^{+}$. Thus,
we will measure excited state absorptions only from the $1B_{2u}^{+}$
state, without any inter-mixing from $1B_{3u}^{+}$ PA spectrum. Moreover,
employing oriented samples, individual energy manifolds ($A_{g}^{-}$/$B_{3g}^{-}$)
can be investigated by using $x$- or $y$-polarized probe beams.
Similarly, one can measure the $1B_{3u}^{+}$ excited state absorption
by tuning the pump energy to $E(1B_{3u}^{+})$. Therefore, by choosing
suitable pump energies and oriented samples, all the four components
of the PA spectrum of each oligomer computed here, can be measured
separately. We would like to mention here that $1B_{2u}^{+}$ and
$1B_{3u}^{+}$ PA spectra are plotted with respect to their energies
obtained in linear optical spectra, reported in Ref. (\cite{sony-acene-lo}).
In naphthalene, using standard parameters we calculated $1B_{2u}^{+}$
state at 4.45 eV and $1B_{3u}^{+}$ state at 5.99 eV, while screened
parameter values for these states were obtained to be 4.51 eV and
5.30 eV, respectively. In case of anthracene, standard parameters
predicted $E(1B_{2u}^{+})=$3.66 eV and $E(1B_{3u}^{+})=$5.34 eV,
on the other hand screened parameters calculated $1B_{2u}^{+}$ and
$1B_{3u}^{+}$ states at 3.55 eV and 4.64 eV, respectively. The energies
reported in tables (\textit{{cf.}} Ref. \cite{epaps}) are the excitation
energies of those states calculated with respect to the $1A_{g}^{-}$
ground state.

We first present the lower end of the spectrum, which is that region
where experimental information about various excited states is available
through two photon spectroscopy based experiments.\cite{bergman-2A,bergman-3A,dick-2A,dick-3A,mikami-2A}
Although, the energies of these states are not so low in terms of
their absolute values, but these states will appear in the low energy
region of the PA spectrum. In absolute terms, the states for naphthalene
and anthracene in this region are up to 6.5 eV and 5.2 eV, respectively,
while in the PA spectra, they will appear in the range of 0--2 eV
for naphthalene, and 0--1.6 eV for anthracene. Thereafter, we compare
our results with the experimental data and calculations of other authors.
Next, we present the discussion for the higher energy region of the
spectrum which so far has not been probed, either experimentally,
or theoretically. While discussing the high energy features we have
to keep in mind that the ionization potentials of naphthalene and
anthracene are 8.1 eV and 7.4 eV, respectively.

\subsection{Low energy region of the PA spectrum }

\begin{figure*}
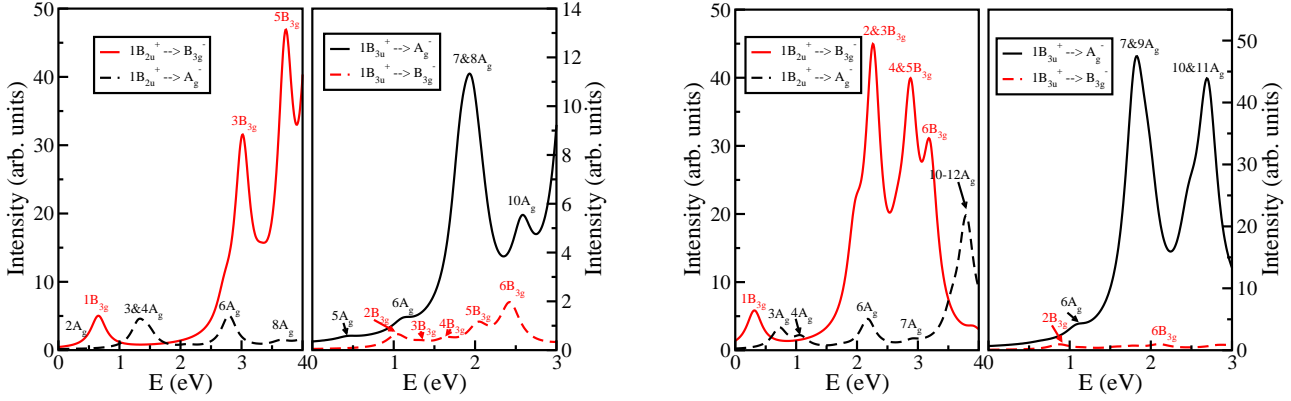

\hspace{-0.7cm}\includegraphics[width=8cm]{fig4a}\hspace{1.0cm}\includegraphics[width=8cm]{fig4b}

\caption{(color online) Photo-induced spectra of anthracene computed using
standard parameters (left panel) and screened parameters (right panel)
in the PPP model Hamiltonian. The left side of both the spectrum depicts
the PA from the $1B_{2u}^{+}$ state, while right side shows it from
the $1B_{3u}^{+}$ state. The $x$/$y$-polarized transitions are
shown using solid/dashed curves. \label{fig:acene3_pa}}

\end{figure*}

The first general aspect of the calculation is the relative location
of the $2A_{g}^{-}$ state of these oligomers with respect to their
$1B_{2u}^{+}$ state. Experimentally speaking both naphthalene and
anthracene exhibit fluorescence, suggesting that $1B_{2u}^{+}$ state
is lower than $2A_{g}^{-}$ state. In our standard parameter calculations
we obtain this ordering of states, while in screened parameter calculation
$2A_{g}^{-}$ state is obtained below $1B_{2u}^{+}$ state. As a result,
$2A_{g}^{-}$ state does not appear in the $1B_{2u}^{+}$ PA spectrum,
computed using the screened parameters, while in standard parameter
spectrum it appears as the first (but very weak) feature in both the
molecules (\emph{cf.} left panel of Fig. (\ref{fig:acene2_pa}) and
Fig. (\ref{fig:acene3_pa})). After the $2A_{g}^{-}$~peak, the next
important feature in the spectra corresponds to the $1B_{3g}^{-}$~state,
which is relatively more intense than the $2A_{g}^{-}$ feature. The
intensity of the $1B_{3g}^{-}$~peak increases, while that of $2A_{g}^{-}$
decreases, with the increase in length of molecule, irrespective of
the parameters used. Both features occur due to transition from the
$1B_{2u}^{+}$~state. Turning over to the $1B_{3u}^{+}$ PA spectra,
we notice that the low energy region gets very faint contributions
from some of the states which contribute to the intermediate energy
region of $1B_{2u}^{+}$ PA spectra.

Now we compare our results for naphthalene and anthracene with the
experiments, and the works of other authors in a quantitative manner.
For both the oligomers, this comparison is summarized in Table \ref{tab:comparion}.
For the case of naphthalene, our results obtained using the standard
parameters are in very good agreement with the available experimental
and other theoretical results, while screened parameters underestimate
the experimental findings (\emph{cf.} Table \ref{tab:comparion}).
Our standard parameter calculations predict a very weak peak corresponding
to $2A_{g}^{-}$~state at 4.91 eV, while $1B_{3g}^{-}$, $3A_{g}^{-}$,
and $4A_{g}^{-}$~states are obtained at 5.35 eV, 5.67 eV, and 6.37
eV, respectively. Upon comparing our results to experiments, we find
that the measured energy\citet{dick-2A,bergman-2A,mikami-2A} of the
$2A_{g}^{-}$ \& $3A_{g}^{-}$ states ($\approx$5.50 eV \& 6.05 eV)
compares very well with our computed energy for $3A_{g}^{-}$ and
$4A_{g}^{-}$ states (5.67 eV \& 6.37 eV). Thus, the measured $nA_{g}^{-}$~state
consequently maps to $(n+1)A_{g}^{-}$ state of our work. We strongly
believe that the $2A_{g}^{-}$ state located around 4.91 eV ($\approx$39601
cm$^{\text{-1}}$) has been missed by the experimentalists due to
its very weak intensity. This, in our opinion calls for further experimental
investigations. While comparing theory and experiments for the anthracene,
a similar mismatch in the labeling of the $A_{g}^{-}$ states was
pointed out by Ramasesha \emph{et al}.\citet{ramasesha-3A} On the
other hand, there is no ambiguity for the $1B_{3g}^{-}$ state, which
is calculated at 5.35 eV by us and measured around 5.20 eV, using
two photon spectroscopy (TPS).\citet{bergman-2A,dick-2A,mikami-2A}
Thus, the disagreement between our theory and experiments is no more
than $5\%$. When this $nA_{g}^{-}$--$(n+1)A_{g}^{-}$ mismatch is
accounted for, we obtain very good agreement between our calculations
and those of Hohlneicher and Dick,\citet{hohlneicher-2A-3A} who reported
$1B_{3g}^{-}$, $2A_{g}^{-}$, and $3A_{g}^{-}$ states at 5.31 eV,
5.45 eV, and 5.91 eV using SDCI/M approach, and at 5.37 eV, 5.70 eV,
and 6.12 eV using SDCI/P approach. Our results also agree well with
the CNDO/S-CI calculations of Swiderek \textit{et al.,}\textit{\emph{\citet{Swiderek1990}}}
who computed $1B_{3g}^{-}$, $2A_{g}^{-}$, and $3A_{g}^{-}$ states
at 5.37 eV, 5.61 eV, and 5.76 eV, respectively. On comparing our results
with the MRMP and CASSCF calculations of Hashimoto \emph{et al.,}\citet{hashimoto-2A}
we found that their CASSCF energies are much higher than ours, as
well as compared to experimental energies, while MRMP results are
comparable. Similarly, SCI results of Pariser\citet{pariser-2A-5A}
predict much higher energies differences as compared to our values.

In anthracene, the standard parameters predict $2A_{g}^{-}$, $1B_{3g}^{-}$,
$3A_{g}^{-}$, and $4A_{g}^{-}$~states at 3.89 eV, 4.31 eV, 4.98
eV, and 5.14 eV, respectively, while the screened parameters values
for the $1B_{3g}^{-}$, $3A_{g}^{-}$, and $4A_{g}^{-}$~states are
3.86 eV, 4.28 eV, and 4.61 eV, respectively. In order to compare our
results with the two photon excitation spectrum measured by Dick and
Hohlneicher,\citet{dick-2A} we found that they reported a very weak
intensity peak around 3.64 eV ($\approx$29300 cm$^{\text{-1}}$)
and associated $A_{g}^{-}$ symmetry to it. Although, they did not
assign any state to this feature, yet from our computed PA spectrum
and excitation energy it compares very favorably with calculated $2A_{g}^{-}$
state, a fact also noted by Ramasesha \emph{et al}.\citet{ramasesha-3A}
Therefore, following the lead of Ramasesha \emph{et al}.\citet{ramasesha-3A},
we also compare experimental $nA_{g}$~states with our $(n+1)A_{g}$~states.
This brings our standard parameters results in very good agreement
with the experimental values of Dick and Hohlneicher,who reported
$1B_{3g}^{-}$, $2A_{g}^{-}$, and $3A_{g}^{-}$~states at 4.44 eV,
4.71 eV, and 5.33 eV, respectively.\citet{dick-3A} Prior to Dick
and Hohlneicher\citet{dick-3A}, Bergman and Jortner\citet{bergman-3A}
also measured the two photon spectrum of anthracene molecule in solution,
as well as of crystalline anthracene. As is obvious from Table \ref{tab:comparion}
that our standard parameter values are in good agreement with both
their data. Since screened parameter results consistently underestimate
the experimental values, we believe that they are not applicable for
the case of anthracene as well. As far as comparison with other theoretical
work is concerned, understandably our results are in excellent agreement
with the work of Ramasesha \textit{et al}.,\citet{ramasesha-3A} who
also used the FCI method in conjunction with the PPP model Hamiltonian.
However, we have gone beyond the work of Ramasesha \textit{et al}.\citet{ramasesha-3A}
by also calculating the PA spectra of anthracene. On comparing our
$(n+1)A_{g}$~states with the $nA_{g}$~states predicted by Kawashima
\emph{et al}.,\cite{kawashima-3A-4A} we obtain a very good agreement
with the MRMP calculations. The calculations performed by other authors
using SDCI/M,\citet{hohlneicher-2A-3A} SDCI/P,\citet{hohlneicher-2A-3A}
CASSCF,\citet{kawashima-3A-4A} and SCI,\citet{pariser-2A-5A} approaches
predict the states discussed at relatively larger energies than our
results.

\subsection{Higher energy region of the PA spectrum }

This higher energy region, which includes the excited states in the
range 6.6 eV to 8.1 eV for naphthalene, and 5.3 eV to 7.4 eV for anthracene,
has so far not been experimentally explored. In this region of the
PA spectra highest intensity transitions correspond to the $x$-polarized
photons, in agreement with the fact that the conjugation length of
the systems is in the $x$-direction. Therefore, $B_{3g}^{-}$ states
lead to the most intense peaks in the $1B_{2u}^{+}$ PA spectra, while
$A_{g}^{-}$ states contribute to the main intensity of the $1B_{3u}^{+}$
PA spectra. However, when these two $x$-polarized transitions are
compared to each other, $B_{3g}^{-}$ peaks are found to be more intense
than $A_{g}^{-}$ peaks. Thus, $B_{3g}^{-}$ states which exhibit
strong dipole coupling to $1B_{2u}^{+}$ may end up playing the same
role in the nonlinear properties of these materials which the $mA_{g}^{-}$
state plays for \emph{trans}-polyacetylene.\cite{mazumdar1991,mazumdar1994}

For naphthalene, with standard parameters $2B_{3g}^{-}$ state (6.58
eV) leads to low intensity features (\emph{cf.} Fig. \ref{fig:acene2_pa}),
while the $3B_{3g}^{-}$ state (8.1 eV) close to the ionization threshold
leads to a very intense peak in $1B_{2u}^{+}$ PA spectrum. Similarly,
$6A_{g}^{-}$ (8.06 eV) also located close to the ionization threshold
contributes a very intense peak to the $1B_{3u}^{+}$ PA spectrum
and a weak one to the $1B_{2u}^{+}$ spectrum. Because of its proximity
to the ionization threshold it may be difficult to detect the $6A_{g}^{-}$
~and $3B_{3g}^{-}$ states, but the $2B_{3g}^{+}$ state should certainly
be detectable.

The PA spectrum of anthracene computed using the standard parameters
is much richer compared to that of naphthalene, and offers many more
detectable peaks. For example, in the \textbf{$B_{3g}^{-}$ }manifold,
states $2B_{3g}^{-}$ (6.37 eV), $3B_{3g}^{-}$ (6.68 eV), and $4B_{3g}^{-}$
(6.98 eV), are well below the ionization threshold of 7.4 eV and lead
to significant peaks either in the $1B_{2u}^{+}$ or $1B_{3u}^{+}$
PA spectra. Particularly noteworthy is the strong intensity of the
$3B_{3g}^{-}$ peak in the $1B_{2u}^{+}$ PA spectrum. $5B_{3g}^{-}$
state at 7.39 eV also leads to an intense peak but may be difficult
to detect because of its closeness to the ionization threshold. As
far as the $A_{g}^{-}$ manifold is concerned, $5A_{g}^{-}$ (5.77
eV), $6A_{g}^{-}$ (6.45 eV), and $7A_{g}^{-}$ (7.16 eV) states contribute
significant intensity to the PA spectrum and have not been explored
so far. Reasonablly strong intensities, coupled with the fact that
they are away from the ionization threshold makes them strong candidates
for experimental investigation. The $8A_{g}^{-}$ (7.3 eV) peak which
is also quite intense is close to the ionization threshold and hence
may be difficult to detect.

As far as the screened parameters based results are concerened, for
both the oligomers more features are visible in the same energy range,
because of the fact that the spectrum is red shifted as compared to
the standard parameters. This trend is in agreement with our earlier
work on linear optical absorption in these materials.\cite{sony-acene-lo}
For example, in naphthalene states all the way up to $4B_{3g}^{-}$
and $7A_{g}^{-}$, while for anthracene states up to $6B_{3g}^{-}$
and $11A_{g}^{-}$, are below the ionization threshold of individual
oligomers, and therefore, in principle are detectable. Thus, the experimental
invesitgation of these high energy features can throw light on the
nature of Coulomb interactions in these materials. Smaller number
of peaks below the ionization threshold will indicate that the standard
parameters describe the Coulomb interactions in these systems, while
the larger number of features will suggest the validity of the screened
parameters. 

Regarding wave functions, one can notice from the tables (\textit{{cf.}}
Ref. \cite{epaps}), that single as well as double excitations contribute
equally to various excitated states visible in these spectra. In particular,
the main configuration contributing to the many-particle wave function
of the $2A_{g}^{-}$~state corresponds to a double excitation described
by HOMO ($H$)$\rightarrow$LUMO ($L$); HOMO ($H$)$\rightarrow$LUMO
($L$) ($|H\rightarrow L;H\rightarrow L\left\rangle \right.$), while
single excitation $|H\rightarrow L+2\left\rangle \right.\pm c.c.$
is found as the major contributor to the many-particle wave function
of the $1B_{3g}^{-}$~state. The most intense $B_{3g}^{-}$ feature
for both naphthalene and anthracene computed using the standard as
well as the screened parameter  occurs mainly due to $|H\rightarrow L;H\rightarrow L+1\left\rangle \right.\pm c.c.$
excitation. For the case of $A_{g}^{-}$ peaks, maximum intensity
can be attributed to states which for both the oligomers, and irrespective
of parameters used, are described by $|H\rightarrow L;H-1\rightarrow L+1\left\rangle \right.$
configuration.

\begin{table*}
\caption{Comparison of results of our calculations performed with the standard
(Std.) parameters and the screened (Scd.) parameters with other experimental
and theoretical results for the most important low-lying even parity
states. The energies of various states are given with respect to the
ground state ($1A_{g}^{-}$). In naphthalene and anthracene, using
the standard parameters $1B_{2u}^{+}$ state was obtained at 4.45
eV and 3.66 eV, respectively.\cite{sony-acene-lo} The screened parameters
predicted $1B_{2u}^{+}$ at 4.51 eV for naphthalene, and 3.55 eV for
anthracene.\cite{sony-acene-lo}}

\raggedright{}\label{tab:comparion}\begin{tabular}{|c|>{\centering}p{1cm}|>{\centering}p{1cm}|c||c||c||cc||c||c||c||c||c||c||c}
\hline 
\multicolumn{15}{|c|}{~~~~~~Excitation energy (eV)}\tabularnewline
\multicolumn{15}{|l|}{State~~~Present work~~~~~~~~~~~Experimental~~~~~~~~~~~~~~~~~~~~~~~~~~Other
theoretical}\tabularnewline
\multicolumn{15}{|l|}{~~~~~~~~~~~Std. ~~~~Scd.}\tabularnewline
\multicolumn{15}{|l|}{~~~~~~~~~~para.~~~~para.}\tabularnewline
\hline 
\multicolumn{15}{|c|}{Naphthalene (C$_{\text{10}}$H$_{\text{8}}$)}\tabularnewline
\hline 
$2A_{g}^{-}$  & 4.91  & -  & \multicolumn{4}{c|}{-} & \multicolumn{8}{c|}{4.88\cite{Ramasesha-2}}\tabularnewline
\hline 
$1B_{3g}^{-}$  & 5.35  & 4.77  & \multicolumn{4}{c|}{5.22,\citet{dick-2A} 5.21,\citet{bergman-2A} 5.27\citet{mikami-2A} } & \multicolumn{8}{c|}{5.31$^{\text{a}}$,\citet{hohlneicher-2A-3A} 5.37$^{\text{b}}$,\citet{hohlneicher-2A-3A}
5.37,\citet{Swiderek1990} 5.74$^{\text{c}}$,\citet{hashimoto-2A}
6.63$^{\text{d}}$,\citet{hashimoto-2A} 5.98\citet{pariser-2A-5A}}\tabularnewline
\hline 
$3A_{g}^{-}$  & 5.67  & 5.01  & \multicolumn{4}{c|}{5.52,\citet{dick-2A} 5.45,\citet{bergman-2A} 5.50\citet{mikami-2A}} & \multicolumn{8}{c|}{5.45$^{\text{a}}$,\citet{hohlneicher-2A-3A} 5.70$^{\text{b}}$,\citet{hohlneicher-2A-3A}
5.61,\citet{Swiderek1990} 5.65$^{\text{c}}$,\citet{hashimoto-2A}
5.86$^{\text{d}}$,\citet{hashimoto-2A} 5.73\citet{pariser-2A-5A}}\tabularnewline
\hline 
$4A_{g}^{-}$  & 6.37  & 5.62  & \multicolumn{4}{c|}{6.05\citet{dick-2A}} & \multicolumn{8}{c|}{5.91$^{\text{a}}$,\citet{hohlneicher-2A-3A} 6.12$^{\text{b}}$,\citet{hohlneicher-2A-3A}
5.76,\citet{Swiderek1990} 6.44$^{\text{c}}$,\citet{hashimoto-2A}
7.04$^{\text{d}}$,\citet{hashimoto-2A} 7.36\citet{pariser-2A-5A}}\tabularnewline
\hline 
\multicolumn{15}{|c|}{Anthracene (C$_{\text{14}}$H$_{10}$)}\tabularnewline
\hline 
$2A_{g}^{-}$  & 3.89  & -  & \multicolumn{4}{c|}{3.64$^{\text{e}}$\citet{dick-3A}} & \multicolumn{8}{c|}{3.87\citet{ramasesha-3A}}\tabularnewline
\hline 
$1B_{3g}^{-}$  & 4.31  & 3.86  & \multicolumn{4}{c|}{4.44$^{\text{e}}$,\citet{dick-3A} 4.53$^{\text{f}}$,\citet{bergman-3A}
4.66$^{\text{e}}$,\citet{bergman-3A} } & \multicolumn{8}{c|}{4.31,\citet{ramasesha-3A} 4.57$^{\text{a}}$,\citet{hohlneicher-2A-3A}
4.52$^{\text{b}}$,\citet{hohlneicher-2A-3A} 4.63$^{\text{c}}$,\citet{kawashima-3A-4A}
6.07$^{\text{d}}$,\citet{kawashima-3A-4A} 4.94\citet{pariser-2A-5A}}\tabularnewline
\hline 
$3A_{g}^{-}$  & 4.98  & 4.28  & \multicolumn{4}{c|}{4.71$^{\text{e}}$,\citet{dick-3A} 4.81$^{\text{f}}$,\citet{bergman-3A}
4.96$^{\text{e}}$,\citet{bergman-3A}} & \multicolumn{8}{c|}{4.96,\citet{ramasesha-3A} 4.29$^{\text{a}}$,\citet{hohlneicher-2A-3A}
4.50$^{\text{b}}$,\citet{hohlneicher-2A-3A} 5.03$^{\text{c}}$,\citet{kawashima-3A-4A}
5.42$^{\text{d}}$,\citet{kawashima-3A-4A} 5.00\citet{pariser-2A-5A}}\tabularnewline
\hline 
$4A_{g}^{-}$  & 5.14  & 4.61  & \multicolumn{4}{c|}{5.33$^{\text{e}}$\citet{dick-3A}} & \multicolumn{8}{c|}{5.12,\citet{ramasesha-3A} 5.18$^{\text{a}}$,\citet{hohlneicher-2A-3A}
5.73$^{\text{b}}$,\citet{hohlneicher-2A-3A} 5.28$^{\text{c}}$,\citet{kawashima-3A-4A}
6.57$^{d}$,\citet{kawashima-3A-4A} 6.82\citet{pariser-2A-5A}}\tabularnewline
\hline
\end{tabular}

$^{\text{a}}$SDCI/M calculations ~~~~~~~~~~~~~~~$^{\text{d}}$CASSCF
calculations

$^{\text{b}}$SDCI/P calculations ~~~~~~~~~~~~~~~~$^{\text{e}}$solution
spectrum

$^{\text{c}}$ MRMP calculations ~~~~~~~~~~~~~~~~$^{\text{f}}$crystalline
spectrum 
\end{table*}

\section{Conclusions}

\label{sec:CONCLUSIONS}In this paper, we presented FCI method based
large scale correlated calculations of photo-induced absorption spectrum
of naphthalene and anthracene from their low-lying one-photon states,
$1B_{2u}^{+}$ and $1B_{3u}^{+}$. The aim behind these calculations
was to understand the nature of important two photon states of these
materials and their coupling to one photon states. The spectra were
computed over a broad energy range, so far unexplored in earlier calculations
as well as experiments. The main conclusions of this work are as follows: 
\begin{enumerate}
\item Our most reliable results were obtained using standard Coulomb parameters
in the Hamiltonian, leading to excellent agreement on the energies
of low lying two photon states, available from two photon spectroscopy
(TPS) experiments. 
\item In particular, the $2A_{g}^{-}$ state was calculated to be higher
than the $1B_{2u}^{+}$ state, predicting these materials to exhibit
fluorescence, in agreement with experimental data. However, two photon
spectroscopy based experiments have not been able to locate the $2A_{g}^{-}$
state, possibly due to its very low intensity. Therefore, we encourage
careful TPS or PA based measurements of the $2A_{g}^{-}$ state in
these materials to verify our predictions. 
\item We make specific predictions about the higher energy region of the
PA spectra, where we found  intense features corresponding both to
the $A_{g}^{-}$ and $B_{3g}^{-}$ type states. These states may also
be visible in the TPS experiments in the higher energy region, which
so far has not been explored. We further predict that the relative
intensities of the peaks corresponding to $x$-polarized transitions
is more than the $y$-polarized ones, which can be probed in experiments
performed with the oriented samples. 
\item We believe that it is feasible to resolve various components of the
PA spectra computed here, by tuning the pump beam to different energies
($E(1B_{2u}^{+})$ or $E(1B_{3u}^{+}$)), and by using oriented samples. 
\end{enumerate}
Because of all the reasons mentioned above it is of interest to probe
the PA spectra of these well established materials; therefore, we
urge strong experimental efforts in this direction.

The calculation of TPA spectrum will additionally verify our results,
as the peaks appearing in the PA spectra should also appear in the
TPA spectra of the considered oligomers. Moreover, some of the important
features of the PA spectra may also appear in the third harmonic generation
(THG) spectra, giving more insight into the states governing the nonlinear
optics of these materials. Therefore, in near future we will present
the results of calculations of the TPA and THG spectra of oligo-acenes,
which are currently underway in our group. 
\begin{acknowledgments}
We thank the Department of Science and Technology (DST), Government
of India, for providing financial support for this work under Grant
No. SR/S2/CMP-13/2006. \end{acknowledgments}

\newpage
   \begin{center}
           {\bf \Large SUPPLEMENTARY INFORMATION} 
    \end{center}

Here we present the tables summarizing the results of our FCI calculations
for excited state absorption in naphthalene and anthracene. The data
presented in the tables includes important configurations contributing
to the many-body wave functions of various excited states, their excitation
energies, and transition dipoles connecting them to the $1B_{2u}^{+}$~state
and $1B_{3u}^{+}$~state.

\vspace{1cm}

\subsection{Naphthalene}

\begin{longtable}{|c|c|c|c|c|}

\caption{$A_{g}^{-}$-type excited states contributing to the photo-induced
absorption spectrum of naphthalene, corresponding to transition from
$1B_{2u}^{+}$ (at 4.45 eV) and $1B_{3u}^{+}$ state (at 5.99 eV)
due to the absorption of $y$-polarized and $x$-polarized photons,
respectively, computed using the FCI method coupled with the standard
parameters in the PPP model Hamiltonian. The table includes many particle
wave functions, excitation energies, and dipole matrix elements of
various states with respect to $1B_{2u}^{+}$ and $1B_{3u}^{+}$ states.
`$+c.c.$' indicates that the coefficient of charge conjugate of a
given configuration has the same sign, while `$-c.c.$' implies that
the two coefficients have opposite signs.\label{tab:acene2-1b2u-1b3u-ag}}
\tabularnewline
\hline 
State  & E (eV)  & \multicolumn{1}{c}{Transition } & Dipole (\AA )  & Wave Functions\tabularnewline
\cline{1-4} 
 &  & $y$-component  & $x$-component  & \tabularnewline
\endfirsthead

\multicolumn{3}{c}%
{{\bfseries \tablename\ \thetable{} -- continued from previous page}} \\
\hline 
State  & E (eV)  & \multicolumn{1}{c}{Transition } & Dipole (\AA )  & Wave Functions\tabularnewline
\cline{3-4} 
 &  & $y$-component  & $x$-component  & \tabularnewline
\endhead

 \multicolumn{5}{|r|}{{Continued on next page}} \\ \hline
\endfoot

\endlastfoot

\hline 
$2A_{g}^{-}$  & 4.91  & 0.162  & -  & $|H-2\rightarrow L+1\left\rangle \right.+c.c.(0.4120)$\tabularnewline
\hline 
 &  &  &  & $|H\rightarrow L;H\rightarrow L\left\rangle \right.(0.4047)$\tabularnewline
\hline 
 &  &  &  & $|H\rightarrow L+3\left\rangle \right.+c.c.(0.3555)$\tabularnewline
\hline 
$3A_{g}^{-}$  & 5.67  & 0.147  & -  & $|H\rightarrow L;H-1\rightarrow L+1\left\rangle \right.(0.4062)$\tabularnewline
\hline 
 &  &  &  & $|H-1\rightarrow L+2\left\rangle \right.-c.c.(0.3918)$\tabularnewline
\hline 
 &  &  &  & $|H\rightarrow L+3\left\rangle \right.+c.c.(0.3253)$\tabularnewline
\hline 
 &  &  &  & $|H\rightarrow L;H\rightarrow L\left\rangle \right.(0.2196)$\tabularnewline
\hline 
$4A_{g}^{-}$  & 6.37  & 0.697  & 0.242  & $|H\rightarrow L;H\rightarrow L\left\rangle \right.(0.6117)$\tabularnewline
\hline 
 &  &  &  & $|H\rightarrow L;H-2\rightarrow L+2\left\rangle \right.(0.2732)$\tabularnewline
\hline 
 &  &  &  & $|H\rightarrow L;H-1\rightarrow L+1\left\rangle \right.(0.2556)$\tabularnewline
\hline 
 &  &  &  & $|H\rightarrow L+2;H\rightarrow L+2\left\rangle \right.+c.c.(0.2537)$\tabularnewline
\hline 
$6A_{g}^{-}$  & 8.06  & 0.300  & 0.734  & $|H\rightarrow L;H-1\rightarrow L+1\left\rangle \right.(0.4981)$\tabularnewline
\hline 
 &  &  &  & $|H\rightarrow L;H-2\rightarrow L+2\left\rangle \right.(0.3886)$\tabularnewline
\hline 
 &  &  &  & $|H\rightarrow L+2;H\rightarrow L+2\left\rangle \right.+c.c.(0.2912)$\tabularnewline
\hline 
 &  &  &  & $|H-2\rightarrow L+4\left\rangle \right.+c.c.(0.2143)$\tabularnewline
\hline 
\end{longtable}

\begin{longtable}{|c|c|c|c|c|}

\caption{$B_{3g}^{-}$-type excited states contributing to the photo-induced
absorption spectrum of naphthalene, corresponding to transition from
$1B_{2u}^{+}$ (at 4.45 eV) and $1B_{3u}^{+}$ state (at 5.99 eV)
due to the absorption of $x$-polarized and $y$-polarized photons,
respectively, computed using the FCI method coupled with the standard
parameters in the PPP model Hamiltonian. The table includes many particle
wave functions, excitation energies, and dipole matrix elements of
various states with respect to $1B_{2u}^{+}$ and $1B_{3u}^{+}$ states.
`$+c.c.$' indicates that the coefficient of charge conjugate of a
given configuration has the same sign, while `$-c.c.$' implies that
the two coefficients have opposite signs.\label{tab:acene2-1b2u-1b3u-b3g}}
\tabularnewline
\hline 
State  & E (eV)  & \multicolumn{1}{c|}{Transition } & Dipole (\AA )  & Wave Functions\tabularnewline
\cline{3-4} 
 &  & $x$-component  & $y$-component  & \tabularnewline
\endfirsthead

\multicolumn{3}{c}%
{{\bfseries \tablename\ \thetable{} -- continued from previous page}} \\
\hline 
State  & E (eV)  & \multicolumn{1}{c}{Transition } & Dipole (\AA )  & Wave Functions\tabularnewline
\cline{3-4} 
 &  & $x$-component  & $y$-component  & \tabularnewline
\endhead

 \multicolumn{5}{|r|}{{Continued on next page}} \\ \hline
\endfoot

\endlastfoot

\hline 
$1B_{3g}^{-}$  & 5.35  & 0.672  & -  & $|H\rightarrow L+2\left\rangle \right.+c.c.(0.5469)$\tabularnewline
\hline 
 &  &  &  & $|H\rightarrow L;H\rightarrow L+1\left\rangle \right.+c.c.(0.2841)$\tabularnewline
\hline 
$2B_{3g}^{-}$  & 6.58  & 0.154  & 0.219  & $|H-1\rightarrow L+3\left\rangle \right.+c.c.(0.4828)$\tabularnewline
\hline 
 &  &  &  & $|H-1\rightarrow L+1;H\rightarrow L+1\left\rangle \right.+c.c.(0.3665)$\tabularnewline
\hline 
$3B_{3g}^{-}$  & 8.12  & 1.271  & 0.255  & $|H\rightarrow L;H\rightarrow L+1\left\rangle \right.+c.c.(0.4081)$\tabularnewline
\hline 
 &  &  &  & $|H\rightarrow L+2\left\rangle \right.+c.c.(0.2349)$\tabularnewline
\hline 
 &  &  &  & $|H\rightarrow L;H\rightarrow L+4\left\rangle \right.+c.c.(0.2335)$\tabularnewline
\hline 
 &  &  &  & $|H-3\rightarrow L;H\rightarrow L+2\left\rangle \right.+c.c.(0.2256)$\tabularnewline
\hline 
 &  &  &  & $|H-2\rightarrow L\left\rangle \right.+c.c.(0.2128)$\tabularnewline
\hline 
\end{longtable}

\begin{longtable}{|c|c|c|c|c|}

\caption{$A_{g}^{-}$-type excited states contributing to the photo-induced
absorption spectrum of naphthalene, corresponding to transition from
$1B_{2u}^{+}$ (at 4.51 eV) and $1B_{3u}^{+}$ state (at 5.30 eV)
due to the absorption of $y$-polarized and $x$-polarized photons,
respectively, computed using the FCI method coupled with the screened
parameters in the PPP model Hamiltonian. The table includes many particle
wave functions, excitation energies, and dipole matrix elements of
various states with respect to $1B_{2u}^{+}$ and $1B_{3u}^{+}$ states.
`$+c.c.$' indicates that the coefficient of charge conjugate of a
given configuration has the same sign, while `$-c.c.$' implies that
the two coefficients have opposite signs.\label{tab:acene2-scr-1b2u-1b3u-ag}}
\tabularnewline
\hline 
State  & E (eV)  & \multicolumn{1}{c|}{Transition } & Dipole (\AA )  & Wave Functions\tabularnewline
\cline{3-4} 
 &  & $y$-component  & $x$-component  & \tabularnewline
\endfirsthead

\multicolumn{3}{c}%
{{\bfseries \tablename\ \thetable{} -- continued from previous page}} \\
\hline 
State  & E (eV)  & \multicolumn{1}{c}{Transition } & Dipole (\AA )  & Wave Functions\tabularnewline
\cline{3-4} 
 &  & $y$-component  & $x$-component  & \tabularnewline
\endhead

 \multicolumn{5}{|r|}{{Continued on next page}} \\ \hline
\endfoot

\endlastfoot
 
\hline 
$3A_{g}^{-}$  & 5.01  & 0.355  & -  & $|H\rightarrow L;H-1\rightarrow L+1\left\rangle \right.(0.4022)$\tabularnewline
\hline 
 &  &  &  & $|H\rightarrow L+3\left\rangle \right.-c.c.(0.3582)$\tabularnewline
\hline 
 &  &  &  & $|H-1\rightarrow L+2\left\rangle \right.+c.c.(0.3396)$\tabularnewline
\hline 
$4A_{g}^{-}$  & 5.62  & 0.856  & 0.612  & $|H\rightarrow L;H\rightarrow L\left\rangle \right.(0.5903)$\tabularnewline
\hline 
 &  &  &  & $|H\rightarrow L;H-2\rightarrow L+2\left\rangle \right.(0.2873)$\tabularnewline
\hline 
 &  &  &  & $|H-1\rightarrow L+2\left\rangle \right.+c.c.(0.2594)$\tabularnewline
\hline 
 &  &  &  & $|H\rightarrow L+2;H\rightarrow L+2\left\rangle \right.-c.c.(0.2494)$\tabularnewline
\hline 
$5A_{g}^{-}$  & 7.25  & 0.336  & 1.331  & $|H\rightarrow L;H-1\rightarrow L+1\left\rangle \right.(0.5808)$\tabularnewline
\hline 
 &  &  &  & $|H\rightarrow L+1;H\rightarrow L+1\left\rangle \right.-c.c.(0.3109)$\tabularnewline
\hline 
$6A_{g}^{-}$  &  & -  & 0.872  & $|H\rightarrow L;H-1\rightarrow L+1\left\rangle \right.(0.3260)$\tabularnewline
\hline 
 &  &  &  & $|H-2\rightarrow L+4\left\rangle \right.+c.c.(0.3156)$\tabularnewline
\hline 
 &  &  &  & $|H\rightarrow L;H-2\rightarrow L+2\left\rangle \right.(0.2960)$\tabularnewline
\hline 
 &  &  &  & $|H\rightarrow L+2;H\rightarrow L+2\left\rangle \right.-c.c.(0.2204)$\tabularnewline
\hline 
$7A_{g}^{-}$  & 8.13  & -  & 0.657  & $|H-4\rightarrow L+2\left\rangle \right.+c.c.(0.2293)$\tabularnewline
\hline 
 &  &  &  & $|H\rightarrow L;H-1\rightarrow L+1\left\rangle \right.(0.2262)$\tabularnewline
\hline 
$8A_{g}^{-}$  & 8.48  & 0.978  & 0.817  & $|H-1\rightarrow L+1;H-1\rightarrow L+1\left\rangle \right.(0.3318)$\tabularnewline
\hline 
 &  &  &  & $|H-2\rightarrow L+2;H-1\rightarrow L+1\left\rangle \right.(0.2959)$\tabularnewline
\hline 
 &  &  &  & $|H\rightarrow L;H-1\rightarrow L+1\left\rangle \right.(0.4022)$\tabularnewline
\hline 
 &  &  &  & $|H\rightarrow L+3\left\rangle \right.-c.c.(0.2825)$\tabularnewline
\hline 
\end{longtable}

\begin{longtable}{|c|c|c|c|c|}

\caption{$B_{3g}^{-}$-type excited states contributing to the photo-induced
absorption spectrum of naphthalene, corresponding to transition from
$1B_{2u}^{+}$ (at 4.51 eV) and $1B_{3u}^{+}$ (at 5.30 eV) states
due to the absorption of $x$-polarized and $y$-polarized photons,
respectively, computed using the FCI method coupled with the screened
parameters in the PPP model Hamiltonian. The table includes many particle
wave functions, excitation energies, and dipole matrix elements of
various states with respect to $1B_{2u}^{+}$ and $1B_{3u}^{+}$ states.
`$+c.c.$' indicates that the coefficient of charge conjugate of a
given configuration has the same sign, while `$-c.c.$' implies that
the two coefficients have opposite signs.\label{tab:acene2-scr-1b2u-1b3u-b3g}}
\tabularnewline
\hline 
State  & E (eV)  & \multicolumn{1}{c|}{Transition } & Dipole (\AA )  & Wave Functions\tabularnewline
\cline{3-4} 
 &  & $x$-component  & $y$-component  & \tabularnewline
\endfirsthead

\multicolumn{3}{c}%
{{\bfseries \tablename\ \thetable{} -- continued from previous page}} \\
\hline 
State  & E (eV)  & \multicolumn{1}{c}{Transition } & Dipole (\AA )  & Wave Functions\tabularnewline
\cline{3-4} 
 &  & $x$-component  & $y$-component  & \tabularnewline
\endhead

 \multicolumn{5}{|r|}{{Continued on next page}} \\ \hline
\endfoot

\endlastfoot

\hline 
$1B_{3g}^{-}$  & 4.77  & 1.119  & -  & $|H\rightarrow L+2\left\rangle \right.+c.c.(0.5519)$\tabularnewline
\hline 
 &  &  &  & $|H\rightarrow L;H\rightarrow L+1\left\rangle \right.-c.c.(0.2247)$\tabularnewline
\hline 
$2B_{3g}^{-}$  & 5.75  & -  & 0.263  & $|H-1\rightarrow L+3\left\rangle \right.+c.c.(0.4766)$\tabularnewline
\hline 
 &  &  &  & $|H-1\rightarrow L+1;H\rightarrow L+1\left\rangle \right.+c.c.(0.3324)$\tabularnewline
\hline 
$3B_{3g}^{-}$  & 7.09  & 1.557  & 0.259  & $|H\rightarrow L;H\rightarrow L+1\left\rangle \right.-c.c.(0.4407)$\tabularnewline
\hline 
 &  &  &  & $|H\rightarrow L+2\left\rangle \right.-c.c.(0.2124)$\tabularnewline
\hline 
 &  &  &  & $|H\rightarrow L+2;H\rightarrow L+3\left\rangle \right.+c.c.(0.2044)$\tabularnewline
\hline 
$4B_{3g}^{-}$  & 7.94  & 1.194  & 0.159  & $|H\rightarrow L+2;H\rightarrow L+3\left\rangle \right.+c.c.(0.2738)$\tabularnewline
\hline 
 &  &  &  & $|H\rightarrow L;H\rightarrow L+4\left\rangle \right.-c.c.(0.2424)$\tabularnewline
\hline 
 &  &  &  & $|H\rightarrow L;H\rightarrow L+1\left\rangle \right.+c.c.(0.2418)$\tabularnewline
\hline 
 &  &  &  & $|H\rightarrow L+2\left\rangle \right.+c.c.(0.2089)$\tabularnewline
\hline 
$5B_{3g}^{-}$  & 8.39  & 0.267  & 0.194  & $|H-4\rightarrow L+3\left\rangle \right.+c.c.(0.3131)$\tabularnewline
\hline 
 &  &  &  & $|H\rightarrow L+3;H-1\rightarrow L+3\left\rangle \right.-c.c.(0.2058)$\tabularnewline
\hline 
\end{longtable}

\newpage

\subsection{Anthracene}

\begin{longtable}{|c|c|c|c|c|}

\caption{$A_{g}^{-}$-type excited states contributing to the photo-induced
absorption spectrum of anthracene, corresponding to transition from
$1B_{2u}^{+}$ (at 3.66 eV) and $1B_{3u}^{+}$ state (at 5.34 eV)
due to the absorption of $y$-polarized and $x$-polarized photons,
respectively, computed using the FCI method coupled with the standard
parameters in the PPP model Hamiltonian. The table includes many particle
wave functions, excitation energies, and dipole matrix elements of
various states with respect to $1B_{2u}^{+}$ and $1B_{3u}^{+}$ states.
`$+c.c.$' indicates that the coefficient of charge conjugate of a
given configuration has the same sign, while `$-c.c.$' implies that
the two coefficients have opposite signs.\label{tab:acene3-1b2u-1b3u-ag}}
\tabularnewline
\hline 
State  & E (eV)  & \multicolumn{1}{c|}{Transition } & Dipole (\AA )  & Wave Functions\tabularnewline
\cline{3-4} 
 &  & $y$-component  & $x$-component  & \tabularnewline
 
 \endfirsthead

\multicolumn{3}{c}%
{{\bfseries \tablename\ \thetable{} -- continued from previous page}} \\
\hline 
State  & E (eV)  & \multicolumn{1}{c}{Transition } & Dipole (\AA )  & Wave Functions\tabularnewline
\cline{3-4} 
 &  & $y$-component  & $x$-component  & \tabularnewline
\endhead

 \multicolumn{5}{|r|}{{Continued on next page}} \\ \hline
\endfoot

\endlastfoot 

\hline 
$2A_{g}^{-}$  & 3.89  & 0.150  & -  & $|H\rightarrow L;H\rightarrow L\left\rangle \right.(0.5639)$\tabularnewline
\hline 
 &  &  &  & $|H\rightarrow L+3\left\rangle \right.+c.c.(0.3372)$\tabularnewline
\hline 
 &  &  &  & $|H-1\rightarrow L+2\left\rangle \right.+c.c.(0.2476)$\tabularnewline
\hline 
$3A_{g}^{-}$  & 4.98  & 0.635  & -  & $|H\rightarrow L+3\left\rangle \right.-c.c.(0.3502)$\tabularnewline
\hline 
 &  &  &  & $|H\rightarrow L;H-1\rightarrow L+1\left\rangle \right.(0.3331)$\tabularnewline
\hline 
 &  &  &  & $|H\rightarrow L;H\rightarrow L\left\rangle \right.(0.3015)$\tabularnewline
\hline 
 &  &  &  & $|H\rightarrow L;H-2\rightarrow L+2\left\rangle \right.(0.2562)$\tabularnewline
\hline 
$4A_{g}^{-}$  & 5.14  & 0.434  & -  & $|H-1\rightarrow L+2\left\rangle \right.+c.c.(0.4913)$\tabularnewline
\hline 
 &  &  &  & $|H\rightarrow L;H\rightarrow L\left\rangle \right.(0.3685)$\tabularnewline
\hline 
 &  &  &  & $|H\rightarrow L;H-1\rightarrow L+1\left\rangle \right.(0.2605)$\tabularnewline
\hline 
$5A_{g}^{-}$  & 5.77  & 0.147  & 0.201  & $|H\rightarrow L;H\rightarrow L+4\left\rangle \right.-c.c.(0.3709)$\tabularnewline
\hline 
 &  &  &  & $|H\rightarrow L+6\left\rangle \right.+c.c.(0.3412)$\tabularnewline
\hline 
 &  &  &  & $|H\rightarrow L;H-2\rightarrow L+2\left\rangle \right.(0.3095)$\tabularnewline
\hline 
$6A_{g}^{-}$  & 6.45  & 0.304  & 0.233  & $|H\rightarrow L;H-2\rightarrow L+2\left\rangle \right.(0.3396)$\tabularnewline
\hline 
 &  &  &  & $|H\rightarrow L+2;H\rightarrow L+2\left\rangle \right.-c.c.(0.3212)$\tabularnewline
\hline 
 &  &  &  & $|H\rightarrow L;H-1\rightarrow L+1\left\rangle \right.(0.2204)$\tabularnewline
\hline 
$7A_{g}^{-}$  & 7.16  & -  & 0.627  & $|H\rightarrow L+1;H\rightarrow L+1\left\rangle \right.+c.c.(4153)$\tabularnewline
\hline 
 &  &  &  & $|H\rightarrow L;H-1\rightarrow L+1\left\rangle \right.(0.4150)$\tabularnewline
\hline 
$8A_{g}^{-}$  & 7.30  & 0.155  & 0.691  & $|H\rightarrow L;H-1\rightarrow L+1\left\rangle \right.(0.3325)$\tabularnewline
\hline 
 &  &  &  & $|H-4\rightarrow L+3\left\rangle \right.+c.c.(0.2810)$\tabularnewline
\hline 
 &  &  &  & $|H\rightarrow L+2;H-1\rightarrow L+3\left\rangle \right.+c.c.(2137)$\tabularnewline
\hline 
 &  &  &  & $|H\rightarrow L;H-4\rightarrow L+4\left\rangle \right.(0.2027)$\tabularnewline
\hline 
$10A_{g}^{-}$  & 7.92  & 0.348  & 0.403  & $|H\rightarrow L;H-1\rightarrow L+1\left\rangle \right.(0.2848)$\tabularnewline
\hline 
 &  &  &  & $|H\rightarrow L;H-2\rightarrow L+2\left\rangle \right.(0.2178)$\tabularnewline
\hline 
\end{longtable}

\begin{longtable}{|c|c|c|c|c|}

\caption{$B_{3g}^{-}$-type excited states contributing to the photo-induced
absorption spectrum of anthracene, corresponding to transition from
$1B_{2u}^{+}$ (at 3.66 eV) and $1B_{3u}^{+}$ state (at 5.34 eV)
due to the absorption of $x$-polarized and $y$-polarized photons,
respectively, computed using the FCI method coupled with the standard
parameters in the PPP model Hamiltonian. The table includes many particle
wave functions, excitation energies, and dipole matrix elements of
various states with respect to $1B_{2u}^{+}$ and $1B_{3u}^{+}$ states.
`$+c.c.$' indicates that the coefficient of charge conjugate of a
given configuration has the same sign, while `$-c.c.$' implies that
the two coefficients have opposite signs.\label{tab:acene3-1b2u-1b3u-b3g} }
\tabularnewline
\hline 
State  & E (eV)  & \multicolumn{1}{c|}{Transition } & Dipole (\AA )  & Wave Functions\tabularnewline
\cline{3-4} 
 &  & $x$-component  & $y$-component  & \tabularnewline
\endfirsthead

\multicolumn{3}{c}%
{{\bfseries \tablename\ \thetable{} -- continued from previous page}} \\
\hline 
State  & E (eV)  & \multicolumn{1}{c}{Transition } & Dipole (\AA )  & Wave Functions\tabularnewline
\cline{3-4} 
 &  & $x$-component  & $y$-component  & \tabularnewline
\endhead

 \multicolumn{5}{|r|}{{Continued on next page}} \\ \hline
\endfoot

\endlastfoot

\hline 
$1B_{3g}^{-}$  & 4.31  & 1.037  & -  & $|H\rightarrow L+2\left\rangle \right.+c.c.(0.5226)$\tabularnewline
\hline 
 &  &  &  & $|H\rightarrow L;H\rightarrow L+1\left\rangle \right.+c.c.(0.2808)$\tabularnewline
\hline 
$2B_{3g}^{-}$  & 6.37  & -  & 0.278  & $|H\rightarrow L+2\left\rangle \right.-c.c.(0.6268)$\tabularnewline
\hline 
$3B_{3g}^{-}$  & 6.68  & 1.168  & 0.126  & $|H-1\rightarrow L+3\left\rangle \right.-c.c.(0.3000)$\tabularnewline
\hline 
 &  &  &  & $|H\rightarrow L;H\rightarrow L+1\left\rangle \right.+c.c.(0.2970)$\tabularnewline
\hline 
$4B_{3g}^{-}$  & 6.98  & -  & 0.155  & $|H-4\rightarrow L+2\left\rangle \right.+c.c.(0.3138)$\tabularnewline
\hline 
 &  &  &  & $|H\rightarrow L+1;H\rightarrow L+4\left\rangle \right.+c.c.(0.2492)$\tabularnewline
\hline 
$5B_{3g}^{-}$  & 7.39  & 1.253  & 0.245  & $|H-2\rightarrow L+2;H\rightarrow L+1\left\rangle \right.+c.c.(0.2496)$\tabularnewline
\hline 
 &  &  &  & $|H\rightarrow L;H\rightarrow L+1\left\rangle \right.-c.c.(0.2449)$\tabularnewline
\hline 
 &  &  &  & $|H\rightarrow L;H-1\rightarrow L+4\left\rangle \right.+c.c.(0.2101)$\tabularnewline
\hline 
 &  &  &  & $|H-4\rightarrow L+2\left\rangle \right.+c.c.(0.2009)$\tabularnewline
\hline 
$6B_{3g}^{-}$  & 7.77  & 1.313  & 0.329  & $|H\rightarrow L;H\rightarrow L+1\left\rangle \right.-c.c.(0.2409)$\tabularnewline
\hline 
 &  &  &  & $|H\rightarrow L+2;H\rightarrow L+3\left\rangle \right.+c.c.(0.2293)$\tabularnewline
\hline 
\end{longtable}

\begin{longtable}{|c|c|c|c|c|}

\caption{$A_{g}^{-}$-type excited states contributing to the photo-induced
absorption spectrum of anthracene, corresponding to transition from
$1B_{2u}^{+}$ (at 3.55 eV) and $1B_{3u}^{+}$ state (at 4.64 eV)
due to the absorption of $y$-polarized and $x$-polarized photons,
respectively, computed using the FCI method coupled with the screened
parameters in the PPP model Hamiltonian. The table includes many particle
wave functions, excitation energies, and dipole matrix elements of
various states with respect to $1B_{2u}^{+}$ and $1B_{3u}^{+}$ states.
`$+c.c.$' indicates that the coefficient of charge conjugate of a
given configuration has the same sign, while `$-c.c.$' implies that
the two coefficients have opposite signs.\label{tab:acene3-scr-1b2u-1b3u-ag}}
\tabularnewline
\hline 
State  & E (eV)  & \multicolumn{1}{c|}{Transition } & Dipole (\AA )  & Wave Functions\tabularnewline
\cline{3-4} 
 &  & $y$-component  & $x$-component  & \tabularnewline
\endfirsthead

\multicolumn{3}{c}%
{{\bfseries \tablename\ \thetable{} -- continued from previous page}} \\
\hline 
State  & E (eV)  & \multicolumn{1}{c}{Transition } & Dipole (\AA )  & Wave Functions\tabularnewline
\cline{3-4} 
 &  & $y$-component  & $x$-component  & \tabularnewline
\endhead

 \multicolumn{5}{|r|}{{Continued on next page}} \\ \hline
\endfoot

\endlastfoot

\hline 
$3A_{g}^{-}$  & 4.27  & 0.782  & -  & $|H\rightarrow L+3\left\rangle \right.+c.c.(0.3743)$\tabularnewline
\hline 
 &  &  &  & $|H\rightarrow L;H-1\rightarrow L+1\left\rangle \right.(0.3083)$\tabularnewline
\hline 
 &  &  &  & $|H\rightarrow L;H\rightarrow L\left\rangle \right.(0.2538)$\tabularnewline
\hline 
 &  &  &  & $|H\rightarrow L;H-2\rightarrow L+2\left\rangle \right.(0.2284)$\tabularnewline
\hline 
$4A_{g}^{-}$  & 4.62  & 0.461  & -  & $|H\rightarrow L+3\left\rangle \right.+c.c.(0.4822)$\tabularnewline
\hline 
 &  &  &  & $|H\rightarrow L;H\rightarrow L\left\rangle \right.(0.3492)$\tabularnewline
\hline 
 &  &  &  & $|H\rightarrow L;H-1\rightarrow L+1\left\rangle \right.(0.2122)$\tabularnewline
\hline 
$5A_{g}^{-}$  & 5.28  & 0.139  & -  & $|H\rightarrow L;H\rightarrow L+4\left\rangle \right.-c.c.(0.3492)$\tabularnewline
\hline 
 &  &  &  & $|H\rightarrow L+6\left\rangle \right.-c.c.(0.3427)$\tabularnewline
\hline 
 &  &  &  & $|H\rightarrow L;H-2\rightarrow L+2\left\rangle \right.(0.2860)$\tabularnewline
\hline 
$6A_{g}^{-}$  & 5.72  & 0.536  & 0.453  & $|H\rightarrow L;H-2\rightarrow L+2\left\rangle \right.(0.3011)$\tabularnewline
\hline 
 &  &  &  & $|H\rightarrow L+2;H\rightarrow L+2\left\rangle \right.+c.c.(0.2867)$\tabularnewline
\hline 
$7A_{g}^{-}$  & 6.45  & 0.178  & 1.767  & $|H\rightarrow L;H-1\rightarrow L+1\left\rangle \right.(0.5117)$\tabularnewline
\hline 
 &  &  &  & $|H-3\rightarrow L+4\left\rangle \right.-c.c.(0.2266)$\tabularnewline
\hline 
 &  &  &  & $|H\rightarrow L+1;H\rightarrow L+1\left\rangle \right.+c.c.(0.2084)$\tabularnewline
\hline 
$9A_{g}^{-}$  & 6.62  & -  & 1.038  & $|H-4\rightarrow L+3\left\rangle \right.+c.c.(0.2661)$\tabularnewline
\hline 
 &  &  &  & $|H\rightarrow L;H-1\rightarrow L+1\left\rangle \right.(0.2513)$\tabularnewline
\hline 
$10A_{g}^{-}$  & 7.11  & 0.373  & 0.403  & $|H\rightarrow L;H-2\rightarrow L+2\left\rangle \right.(0.2783)$\tabularnewline
\hline 
 &  &  &  & $|H\rightarrow L;H-1\rightarrow L+1\left\rangle \right.(0.2491)$\tabularnewline
\hline 
 &  &  &  & $|H-2\rightarrow L+2;H\rightarrow L+4\left\rangle \right.+c.c.(0.2076)$\tabularnewline
\hline 
$11A_{g}^{-}$  & 7.34  & 0.834  & 1.437  & $|H\rightarrow L+3\left\rangle \right.+c.c.(0.2264)$\tabularnewline
\hline 
 &  &  &  & $|H-2\rightarrow L+5\left\rangle \right.+c.c.(0.2144)$\tabularnewline
\hline 
 &  &  &  & $|H\rightarrow L;H-1\rightarrow L+1\left\rangle \right.(0.2128)$\tabularnewline
\hline 
 &  &  &  & $|H\rightarrow L;H\rightarrow L\left\rangle \right.(0.2086)$\tabularnewline
\hline 
$12A_{g}^{-}$  & 7.63  & 0.206  & -  & $|H\rightarrow L+1;H-3\rightarrow L+2\left\rangle \right.+c.c.(0.2416)$\tabularnewline
\hline 
 &  &  &  & $|H-1\rightarrow L+1;H-2\rightarrow L+2\left\rangle \right.(0.2401)$\tabularnewline
\hline 
\end{longtable}

\begin{longtable}{|c|c|c|c|c|}

\caption{$B_{3g}^{-}$-type excited states contributing to the photo-induced
absorption spectrum of anthracene, corresponding to transition from
$1B_{2u}^{+}$ (at 3.55 eV) and $1B_{3u}^{+}$ (at 4.64 eV) states
due to the absorption of $x$-polarized and $y$-polarized photons,
respectively, computed using the FCI method coupled with the screened
parameters in the PPP model Hamiltonian. The table includes many particle
wave functions, excitation energies, and dipole matrix elements of
various states with respect to $1B_{2u}^{+}$ and $1B_{3u}^{+}$ states.
`$+c.c.$' indicates that the coefficient of charge conjugate of a
given configuration has the same sign, while `$-c.c.$' implies that
the two coefficients have opposite signs.\label{tab:acene3-scr-1b2u-1b3u-b3g}}
\tabularnewline
\hline 
State  & E (eV)  & \multicolumn{1}{c|}{Transition } & Dipole (\AA )  & Wave Functions\tabularnewline
\cline{3-4} 
 &  & $x$-component  & $y$-component  & \tabularnewline
\endfirsthead

\multicolumn{3}{c}%
{{\bfseries \tablename\ \thetable{} -- continued from previous page}} \\
\hline 
State  & E (eV)  & \multicolumn{1}{c}{Transition } & Dipole (\AA )  & Wave Functions\tabularnewline
\cline{3-4} 
 &  & $x$-component  & $y$-component  & \tabularnewline
\endhead

 \multicolumn{5}{|r|}{{Continued on next page}} \\ \hline
\endfoot

\endlastfoot

\hline 
$1B_{3g}^{-}$  & 3.86  & 1.593  & -  & $|H\rightarrow L+2\left\rangle \right.+c.c.(0.5375)$\tabularnewline
\hline 
 &  &  &  & $|H\rightarrow L;H\rightarrow L+1\left\rangle \right.-c.c.(0.2162)$\tabularnewline
\hline 
$2B_{3g}^{-}$  & 5.52  & 0.971  & 0.373  & $|H-1\rightarrow L+3\left\rangle \right.+c.c.(0.3189)$\tabularnewline
\hline 
 &  &  &  & $|H-1\rightarrow L+1;H\rightarrow L+1\left\rangle \right.+c.c.(0.2503)$\tabularnewline
\hline 
 &  &  &  & $|H\rightarrow L;H\rightarrow L+1\left\rangle \right.-c.c.(0.2053)$\tabularnewline
\hline 
$3B_{3g}^{-}$  & 5.81  & 1.612  & 0.140  & $|H\rightarrow L;H\rightarrow L+1\left\rangle \right.-c.c.(0.3673)$\tabularnewline
\hline 
 &  &  &  & $|H-1\rightarrow L+3\left\rangle \right.+c.c.(0.3316)$\tabularnewline
\hline 
$4B_{3g}^{-}$  & 6.21  & 0.613  & 0.153  & $|H-4\rightarrow L+2\left\rangle \right.+c.c.(0.4228)$\tabularnewline
\hline 
$5B_{3g}^{-}$  & 6.43  & 1.271  & 0.174  & $|H\rightarrow L;H\rightarrow L+5\left\rangle \right.+c.c.(0.2224)$\tabularnewline
\hline 
 &  &  &  & $|H-2\rightarrow L+2;H\rightarrow L+1\left\rangle \right.+c.c.(0.2040)$\tabularnewline
\hline 
$6B_{3g}^{-}$  & 6.75  & 1.050  & 0.220  & $|H\rightarrow L+2;H\rightarrow L+3\left\rangle \right.+c.c.(0.2451)$\tabularnewline
\hline 
\end{longtable}

\end{document}